\documentclass{article}

\newcommand{\iec}{\mbox{i.\,e.\,}}

\newcommand{\egc}{\mbox{e.\,g.\,}}

\newcommand{\vctr}[1]{\ensuremath{\mathbf{ #1 }}}

\newcommand{\pb}[2]{\ensuremath{\left\{ #1 , #2 \right\} }}

\newcommand{\dr}[1]{\ensuremath{\mathrm{d} #1\,}}
\newcommand{\mc}[1]{\ensuremath{\mathcal{#1}}}

\newcommand{\ddt}{\ensuremath{\frac{\dr{}}{\dr{t}}}}

\newcommand{\pbp}[2]{\ensuremath{\frac{\partial #1}{\partial #2}}}

\newcommand{\edf}{\ensuremath{=_{_{df}}}}

\newcommand{\ket}[1]{\ensuremath{\left|  #1 \right\rangle}}
\newcommand{\bra}[1]{\ensuremath{\left\langle #1 \right|}}
\newcommand{\bk}[2]{\ensuremath{\left\langle #1 | #2 \right\rangle}}
\newcommand{\proj}[2]{\ensuremath{\ket{#1} \bra{#2}}}

\newcommand{\matel}[3]{\ensuremath{\bra{#1} #2 \ket{#3}}}
 
\newcommand{\op}[1]{\ensuremath{\widehat{\textsf{\ensuremath{#1}}}}}

\newcommand{\id}{\op{\mathsf{1}}}

\newcommand{\comm}[2]{\ensuremath{\left[ #1 , #2 \right]}} 
\newcommand{\tr}{\textsf{Tr}}

\newcommand{\be}{\begin{equation}}
\newcommand{\ee}{\end{equation}}
\newcommand{\e}[1]{\mathrm{e}^{#1}}

\usepackage{chicago}
\usepackage{float}
\usepackage{tikz}
\usepackage{tikz-cd}

\newcommand{\mb}[2]{\ensuremath{\left\{ #1 , #2 \right\}_{MB} }}

\begin{document}

\title{Probability and Irreversibility in Modern Statistical Mechanics: Classical and Quantum}
\author{David Wallace}
\date{July 1 2016}
\maketitle

\begin{abstract}
Through extended consideration of two wide classes of case studies --- dilute gases and linear systems --- I explore the ways in which assumptions of probability and irreversibility occur in contemporary statistical mechanics, where the latter is understood as primarily concerned with the derivation of quantitative higher-level equations of motion, and only derivatively with underpinning the equilibrium concept in thermodynamics. I argue that at least in this wide class of examples, (i) irreversibility is introduced through a reasonably well-defined initial-state condition which does not precisely map onto those in the extant philosophical literature; (ii) probability is explicitly required both in the foundations and in the predictions of the theory. I then consider the same examples, as well as the more general context, in the light of quantum mechanics, and demonstrate that while the analysis of irreversibility is largely unaffected by quantum considerations, the notion of statistical-mechanical probability is entirely reduced to quantum-mechanical probability.
\end{abstract}

\section{Introduction: the plurality of dynamics}\label{introduction}

What are the dynamical equations of physics? A first try: 

\begin{quote}
Before the twentieth century, they were the equations of classical mechanics: Hamilton's equations, say,
\be 
\dot{q}^i= \pbp{H}{p_i}\,\,\,\,\,\,\dot{p}_i=-\pbp{H}{q^i}.
\ee
Now we know that they are the equations of quantum mechanics:\footnote{For the purposes of this article I assume that the Schr\"{o}dinger equation is exact in quantum mechanics, and that `wavefunction collapse', whatever its significance, is at any rate not a \emph{dynamical} process. This clearly excludes the GRW theory \cite{grw,pearle,bassighirardireview} and similar dynamical-collapse theories (see \citeN{alberttimechance} for extensive consideration of their statistical mechanics); it includes the Everett interpretation (\citeNP{everett,wallacebook}) and hidden-variable theories like Bohmian mechanics (\citeNP{bohm,bohmhiley,durrincushing}); in \citeN{wallaceorthodoxy} I argue that it includes orthodox quantum mechanics, provided `orthodoxy' is understood via physicists' actual practice.} that is, Schr\"{o}dinger's equation,
\be 
\ddt \ket{\psi}= -\frac{i}{\hbar}\op{H}\ket{\psi}.
\ee

\end{quote}
This answer is misleading, bordering on a category error. Hamilton's equations, and Schr\"{o}dinger's, are not concrete systems of equations but frameworks in which equations may be stated. For any choice of phase space or Hilbert space, and any choice of (classical or quantum) Hamiltonian on that space, we get a set of dynamical equations. In classical mechanics, different choices give us:
\begin{itemize}
\item The Newtonian equations for point particles (better: rigid spheres) interacting under gravity, generalisable to interactions under other potential-based forces;
\item The simple harmonic oscillator equations that describe springs and other vibrating systems;
\item Euler's equations, describing the rotations of rigid bodies;
\item Euler's (other) equations, describing fluid flow in the regime where viscosity can be neglected.
\item (With a little care as to how the Hamiltonian dynamical framework handles them), the field equations of classical electromagnetism and general relativity.
\end{itemize}
In quantum mechanics, we also have
\begin{itemize}
\item The quantum version of the Newtonian equations, applicable to (e.g.) nonrelativistic point particles interacting under some potential;
\item The quantum version of the harmonic oscillator;
\item The quantum field theories of solid-state physics, describing such varied systems as superconductors and vibrating crystals;
\item The quantum field theories of particle physics, in particular the Standard Model, generally thought by physicists to underly pretty much all phenomena in which gravity can be neglected and most in which it cannot.
\end{itemize}
All of these equations are widely used in contemporary physics, and all have been thoroughly confirmed empirically in applications to the systems to which they apply. And this is already something of a puzzle: how is it that physics doesn't just get by with one such system, the \emph{right} such system? Even if that system is far too complicated to solve in practice, that in itself does not guarantee us the existence of other, simpler, equations, applicable in domains where the Right System cannot be applied? Any systematic understanding of physics has to grapple with the plurality of different dynamical equations it uses. Indeed, it must grapple with the plurality of \emph{frameworks}, for the classical equations, too, remain in constant use in contemporary physics.  The problem of `the quantum-classical transition' is poorly understood if it is seen as a transition from one dynamical system (quantum mechanics) to another; it is, rather, a problem of understanding the many concrete classical dynamical equations we apply in a wide variety of different situations.

But even this account greatly understates the plurality of dynamics. Some more examples still:
\begin{itemize}
\item The equations of radioactive, or atomic, decay.
\item The Navier-Stokes equation, describing fluid flow when viscocity \emph{cannot} be neglected.
\item The Langevin equation, a stochastic differential equation describing the behaviour of a large body in a bath of smaller ones, and the related Fokker-Planck equation, describing the evolution of a probability distribution under the Langevin equation.
\item The Vlasov equation, describing the evolution of the particles in a plasma --- or the stars in a cluster or galaxy --- in the so-called ``collisionless'' regime.
\item The Boltzmann equation, describing the evolution of the atoms or molecules in a dilute (but not `collisionless') gas.
\item The Balescu-Lenard equation, describing the evolution of plasmas outside the collisionless regime. 
\item The master equations of chemical dynamics, describing the change in chemical (or nuclear) composition of a fluid when particles of one species can react to form particles of another.
\end{itemize}
It is (as we shall see) a delicate matter which of these equations is classical and which quantum. But at any rate, none of them fit into the dynamical framework of either Hamiltonian classical mechanics or unitary quantum mechanics. In fact, for the most part they have two features that are foreign to both:
\begin{enumerate}
\item They are \emph{probabilistic}, either through being stochastic or, more commonly, by describing the evolution of probability distributions in a way that cannot be reduced to deterministic evolution of individual states. (In the quantum case, they describe the evolution of mixed states in a way that cannot be reduced to deterministic --- much less unitary --- evolution of pure states.)
\item They are \emph{irreversible}: in a sense to be clarified shortly, they build in a clear direction of time, whereas both Hamilton's equations and the Schr\"{o}dinger equation describe dynamics which are in an important sense time-reversal-invariant.
\end{enumerate}
And again, these equations --- and others like them --- are very widely used throughout physics, and very thoroughly tested in their respective domains.

One can imagine a world in which physics is simply a patchwork of overlapping domains each described by its own domain-specific set of equations, with few or no connections between them. (Indeed, Nancy Cartwright~\citeyear{Cartwright1983,cartwrightdappled} makes the case that the actual world is like this.) But the consensus in physics is that in the \emph{actual} world, these various dynamical systems are connected: equation A and equation B describe the same system at different levels of detail, and equation A turns out to be derivable from equation B (perhaps under certain assumptions). Indeed, it is widely held in physics that \emph{pretty much every} dynamical equation can ultimately, in principle, be derived from the Standard Model, albeit sometimes through very long chains of inference via lots of intermediate steps.

I don't think it does great violence to physics usage to say that (non-equilibrium) statistical mechanics just \emph{is} the process of carrying out these constructions, especially (but not only) in the cases where the derived equations are probabilistic and/or irreversible. But in any case, in this paper I want to pursue a characterisation of the \emph{conceptual foundations} of (non-equilibrium) statistical mechanics as concerned with understanding how such constructions can be carried out and what additional assumptions they make. The main conceptual puzzles then concern precisely (1) the probabilistic and (2) the irreversible features of (much) emergent higher-level dynamics; they can be dramatised as a contradiction between two apparent truisms:
\begin{enumerate}
\item Derivation of probabilistic, irreversible higher-level physics from non-probabilistic, time-reversal-invariant lower-level physics is \emph{impossible}: probabilities cannot be derived from non-probabilistic inputs without some explicitly probabilistic assumptions, and more importantly, as a matter of logic time-reversal invariant low-level dynamics cannot allow us to derive time-reversal-noninvariant results.
\item Derivation of probabilistic, irreversible higher-level physics from non-probabilistic, time-reversal-invariant lower-level physics is \emph{routine}: in pretty much all of the examples I give, and more besides, apparent derivations can be found in the textbooks. And these derivations have novel predictive success, both qualitatively (in many cases, such as the Boltzmann, Vlasov and Balescu-Lenard equations, higher-level equations in physics are derived from the lower level and then tested, rather than established phenomenologically and only then derived) and quantitatively (statistical mechanics provides calculational methods to work out the coefficients and parameters in higher-level equations from lower-level inputs, and does so very successfully).
\end{enumerate}
I want to suggest that we can make progress on resolving this apparent paradox through three related strategies:
\begin{enumerate}
\item Focussing on the derivation of quantitative equations of motion, rather than on the qualitative problem of how irreversibility could \emph{possibly} emerge from underlying microphysics. The advantage of the former strategy is that we actually have concrete `derivations' of irreversible, probabilistic microphysics all over modern statistical mechanics; identifying the actual assumptions made in those derivations, and in particular focussing in on those assumptions which have a time-reversal-noninvariant character, is \emph{prima facie} a much more tractable task than trying to work out in the abstract what those assumptions might be.
\item Looking at examples from (relatively) contemporary physics. Insofar as philosophy of statistical mechanics has considered the quantitative features of non-equilibrium statistical mechanics, it has \emph{generally} focussed on Boltzmann's original derivation of the eponymous equation, and of later attempts to improve the mathematical rigor of that equation whilst keeping its general conceptual structure. But contemporary statistical mechanics tends to treat the Boltzmann equation in a different (more explicitly probabilistic) way, as well as incorporating a wide range of additional examples.
\item Engaging fully with quantum as well as classical mechanics. It is common to see, in foundational discussions, the claim that the relevant issues go over from classical to quantum \emph{mutatis mutandis}, so that we can get away with considering only classical physics. \citeN[p.12]{sklarstatmech} is typical: 
\begin{quote}
the particular conceptual problems on which we focus --- the origin and rationale of probability distribution assumptions over initial states, the justification of irreversible kinetic equations, and so on --- appear, for the most part, in similar guise in the development of both the classical and quantum versions of the theory. The hope is that by exploring these issues in the technically simpler classical case, insights will be gained that will carry over to the understanding of the corrected version of the theory.
\end{quote}
 I will argue that this position, while defensible as regards irreversibility, fails radically for probability --- something that could perhaps have been predicted when we recall that quantum theory is itself an inherently probabilistic theory.
\end{enumerate}

In sections \ref{bbgky}--\ref{coulddobetter} I look at modern non-equilibrium statistical mechanics through two detailed classes of example: dilute gases (including the famous Boltzmann equation, where I contrast a modern, probabilistic analysis (section \ref{boltzmann-modern}) with Boltzmann's own approach (section \ref{boltzmann-contrast}) and linear systems (in particular Brownian motion in an oscillator environment, which gives rise to the stochastic Langevin equation and the related Fokker-Planck equation). In both cases, the name of the game is to derive closed-form equations from some coarse-grained description of the system. And two common features of these derivations stand out: explicit and irreducible appeal to the Liouville form of classical mechanics, and a concrete, quantitative condition for irreversibility expressed in terms of an initial-state condition that constrains the residual (\iec, non-coarse-grained) features. 

In sections \ref{mixedstates}--\ref{conclusion} I consider the quantum case, first in general terms (section \ref{mixedstates}), where the analogy between quantum mixed states 
and classical probability distributions turns out to be superficial, and then in the concrete context of the quantum versions of the dilute gas and of Brownian motion --- where ``quantum version'' does not mean ``some new system, analogous to the old'', but ``the same old system, analysed correctly once we remember that quantum mechanics is the correct underlying dynamics''. I conclude that while the problem of irreversibility looks pretty similar classically and quantum-mechanically, the problem of probability is radically transformed: statistical-mechanical probabilities reduce entirely to quantum-mechanical ones.

The physics discussed in this paper is standard and I do not give original references. I have followed \citeN{balescubook} (for dilute gases), \citeN{zehtime} and Zwanzig~\citeyear{Zwanzig1960,Zwanzig1966} for the general linear-systems formalism, \citeN{zwanzigbook} for the Langevin and Fokker-Planck equations, \citeN{pazzurekreview} for the perturbative derivation of the Fokker-Planck equation, \citeN{peres} for the phase-space version of quantum mechanics, and \citeN{liboffkinetic} for the quantum version of the BBGKY hierarchy.

\section{The BBGKY hierarchy and the Vlasov equation}\label{bbgky}

Consider $N$ indistinguishable classical point particles (for very large $N$: $10^6$ at least, perhaps larger), interacting via the Hamiltonian
\begin{eqnarray}
H_N&=&\sum_{1\leq i \leq N}\frac{\vctr{p}_i \cdot \vctr{p}_i}{2m} + \sum_{1\leq i < j \leq N}V(\vctr{q}^i-\vctr{q}^j)
\nonumber
\\
&\equiv & H_1(\vctr{p}_i,\vctr{q}^i) + \sum_{1\leq i < j \leq N}V(\vctr{q}^i-\vctr{q}^j).
\end{eqnarray}
With the right choices of the interaction term $V$, this might describe the particles in  a `classical' gas or plasma, or the stars in a galaxy or cluster. 

A standard move in contemporary statistical mechanics is to consider the evolution of a \emph{probability distribution} $\rho$ over phase space under the Hamiltonian dynamics defined by this Hamiltonian. (Thus probability is introduced explicitly and by hand; I offer no justification for this at this stage, beyond the fact that it is in fact routinely done!) It is well known that any such distribution evolves by the Liouville equation,
\be
\ddt\rho=\pb{H_N}{\rho}=\sum_{1\leq i \leq N}\left( \pbp{H_N}{\vctr{q}^i} \cdot \pbp{\rho}{\vctr{p}_i}- \pbp{H_N}{\vctr{q}^i} \cdot \pbp{\rho}{\vctr{p}_i}\right).
\ee
If we assume the distribution $\rho$ is symmetric under particle interchange (or, equivalently, if we just work with the symmetrised version of $\rho$, which is empirically equivalent), and if we write $x^i$, schematically, for the six coordinates $\vctr{q}^i,\vctr{p}_i$, then we can define, for each $M\leq N$, the $M$-particle marginal probability as
\be
\rho_M = \prod_{M<i\leq N}\int \dr{x}^i \rho(x^1,\ldots x^N).
\ee
$\rho_m$ represents the probability that a randomly-selected $M$-tuple of particles will be found in a given region of their $M$-particle phase space.
We can, iteratively, define $m$-particle correlation functions: the 2-particle correlation function is
\be 
c_2(x^1,x^2)=\rho_2(x^1,x^2)-\rho_1(x^1)\rho_1(x^2),
\ee
the three-particle correlation function is
\begin{eqnarray}
c_3(x^1,x^2,x^3)&=&\rho_3(x^1,x^2,x^3)-\rho_1(x^1)\rho_1(x^2)\rho_1(x^3) \nonumber \\
& - & \rho_2(x^1,x^2)c_2(x^3) - \rho_2(x^1,x^3)c_2(x^2) - \rho_2(x^2,x^3)c_2(x^1)
\end{eqnarray}
and so forth.

The \emph{BBGKY hierarchy} (named for Bogoliubov, Born, Green, Kirkwood and Yvon) is a rewriting of the Liouville equation in terms of the $n$-particle marginals. For each $M$, an equation of motion  can be written down that is \emph{almost} a closed first-order differential equation for $\rho_M$, but which has a term in $\rho_{M+1}$:
\begin{eqnarray}
\ddt{\rho_M}&=&\pb{H_M}{\rho_M} \nonumber \\
&+ &(N\!-\!M)\sum_{1\leq i\leq M}\int \dr{x}^{M+1}\vctr{\nabla}V(\vctr{q}^i-\vctr{q}^{M+1})\cdot \pbp{\rho_{M+1}}{\vctr{p}_i}
\end{eqnarray}
 So the first equation in the hierarchy can be used to determine the evolution of $\rho_1$, but only given the value of $\rho_2$; the latter can in turn be determined from the \emph{second} equation, and so forth. For instance, the first equation in the hierarchy,
\be 
\ddt{\rho}_1 = \pb{H_1}{\rho_1} + (N\!-\!1)\int \dr{x^2} \vctr{\nabla}V(\vctr{q}^1-\vctr{q}^2)\cdot \pbp{\rho_2}{\vctr{p}_2},
\ee
is the equation for free-particle motion plus a correction term linear in the two-particle marginal.

The last equation in the hierarchy is just the $N$-particle Liouville equation, and the full system of equations does not take us beyond that equation. The value of the hierarchy is that it allows us to define various approximations. If we can justify assuming that the $M+1$-particle marginals are negligible, we can truncate the hierarchy at the $M$th equation and obtain a \emph{closed} system of equations for $\rho_M$. The grounds for that assumption will have to be assessed on a case-by-case basis, but crucially, any such assumption does not \emph{in itself} contain any statement of time asymmetry.

For instance, the \emph{collisionless approximation} takes $c_2\simeq 0$, \iec $\rho_2(x^1,x^2)\simeq \rho_1(x^1)\rho_1(x^2)$ (and also $N-1 \simeq N$) and closes the first equation in the hierarchy to give a self-contained equation in $\rho_1$: the \emph{Vlasov equation},
\be
\ddt{\rho_1}(\vctr{q},\vctr{p},t) = \pb{H_1}{\rho_1}(\vctr{q},\vctr{p},t) +N \rho_1(\vctr{q},\vctr{p},t)\int\dr{\vctr{p}'} \dr{\vctr{q}'}\rho_1(\vctr{q}',\vctr{p}',t)\vctr{\nabla}V(\vctr{q}-\vctr{q}'),
\ee
widely used in plasma physics and galactic dynamics.
This approximation is justified by (a) assuming the initial multiparticle correlations are negligible; (b) making additional assumptions about the initial state and the dynamics which jointly entail that negligible multiparticle correlations remain negligible.  Under these assumptions the equation, which on its face describes a one-particle \emph{probability distribution}, can also be taken to describe the \emph{actual distribution} of particles when averaged over regions large compared to (total system volume / N).

I make no claim that the assumptions used to derive the Vlasov equation have been rigorously demonstrated to entail it. But they don't appear to differ from the general run of the mill in mainstream theoretical physics in their level of rigor. And, since they do not distinguish a direction of time, nor does the Vlasov equation: it is probabilistic, but time-reversal-invariant. (It is, however, nonetheless highly non-trivial, despite its superficial resemblance to the one-particle version of the Poisson equation: in particular, the second term in the equation is non-linear in $\rho_1$.) Irreversibility will require a more complicated equation, to which I now turn.

\section{The Boltzmann equation: a modern approach}\label{boltzmann-modern}

Let's relax the collisionless assumption, but only slightly: in the \emph{dilute-gas assumption}, we assume that \emph{three}-particle correlations are negligible. We also assume a two-particle interaction potential $V$ that decreases rapidly with distance, so that the nonlinear term in the Vlasov equation is negligible. The first two equations in the BBGKY hierarchy are now a closed set of equations in $\rho_1$ and $c_2$, which after some manipulation (see \citeN[ch.7]{balescubook} for the details) yields the following (schematically expressed) \emph{integro-differential equation} for $\rho_1$:
\be \label{protoboltzmann}
\ddt{\rho_1}(t) = \{H_1,\rho_1\}  + \int_0^t \dr{\tau} K(\tau)( \rho_1 \otimes \rho_1)(t-\tau)+ \Lambda(t) \cdot c_2(0)
\ee
Here I write $(\rho_1\otimes \rho_1) (x^1,x^2,t)\equiv \rho_1(x^1,t)\rho_1(x^2,t)$, and suppress the dependence of $\rho_1$ and $c_2$ on anything except time. $\Lambda(t)$ and $K(t)$ are time-dependent linear operators whose precise form will not be needed.

This equation holds only for $t>0$, and might appear to describe explicitly time-reversal-noninvariant dynamics. It does not: it has been derived from the dilute-gas approximation without further assumptions and the latter assumption is time-reversal (and time-translation) invariant. The time reverse of the equation holds for $t<0$, and the whole system distinguishes no preferred direction or origin of time.

It can now be demonstrated --- and again, this is a \emph{mathematical} result, the rigor of which can be questioned but which involves no additional physical assumptions --- that $K(\tau) \rho_1 \otimes \rho_1(t-\tau)$ decreases very rapidly with time, so that (for times short compared to the recurrence time $T$ of the system) we can approximate the third term in (\ref{protoboltzmann}) by
\be
\int_0^t \dr{\tau} K(\tau) (\rho_1 \otimes \rho_1)(t-\tau) \simeq \left(\int_0^T \dr{\tau}K(\tau)\right)(\rho_1\otimes \rho_1)(t).
\ee
And in fact the expression on the right hand side is equal to the well-known \emph{Boltzmann collision term} $\kappa[\sigma,\rho_1(t)]$, which depends on $\rho_1(t)$ (quadratically) and on the scattering cross-section $\sigma(\vctr{p}\vctr{p}'\rightarrow \vctr{k}\vctr{k'})$ for two-particle scattering under the interaction potential $V$. (The latter can be calculated by standard methods of scattering theory.)

The equation (\ref{protoboltzmann}) can now be approximated as
\be \label{protoboltzmann2}
\ddt{\rho_1}(t) = \{H_1,\rho_1(t)\}  + \kappa[\sigma,\rho_1(t)] + \Lambda(t) \cdot c_2(0).
\ee
This differs from the full \emph{Boltzmann equation},
\be 
\ddt{\rho_1}(t) = \{H_1,\rho_1(t)\}  + \kappa[\sigma,\rho_1(t)] ,
\ee
only by the final term $\Lambda(t)\cdot c_2(0)$, which is a dependency of the rate of change of $\rho_1$ at time $t$ on the \emph{initial} two-particle correlation function.

In textbooks (see, \egc, \citeN[ch.7]{balescubook}, one can find heuristic arguments to the effect that this term can be expected to be negligible (perhaps after some short `transient' period): indeed, Balescu quotes Prigogine and co-workers as referring to the $\Lambda(t)\cdot c_2(0)$ term as the \emph{destruction term}. But this assumption needs to be treated cautiously: the Boltzmann equation is time-reversal noninvariant, and so the assumption that this term vanishes must in some way build in assumptions that break the time symmetries of the problem. And indeed, the heuristic arguments would fail if a precisely arranged pattern of delicate correlations were present at time 0: they rely on certain assumptions of genericity about those correlations.

Following \citeN{wallacelogic}, let's call an initial two-particle correlation function (and, by extension, an initial probability distribution satisfying our other assumptions) \emph{forward compatible} if $\Lambda(t)\cdot c_2(0)$ is negligible for at least $0<t \ll T$. Then we have established that systems whose initial state is forward compatible will obey the Boltzmann equation. 

How are we to think about this assumption? I can see three options:
\begin{enumerate}
\item At one extreme, we could simply posit that a system's state is forward compatible, and derive from this that it obeys the Boltzmann equation. This is dangerously close to circularity: to say that a state is forward compatible is close to saying just \emph{that} it obeys the Boltzmann equation. The formal machinery developed here allows us to sidestep that circularity: to say that a correlation function satisfies $\Lambda(t)\cdot c_2(0)\simeq 0$ is equivalent to saying that its forward time evolution satisfies the Boltzmann equation only given significant, non-trivial mathematical work. But we are still left with little clarity as to what this obscure mathematical expression means, and in particular, we fail to connect to the heuristic arguments that this term can in some sense be `expected' to vanish for `reasonable' choices of $c_2(0)$.
\item At the other extreme, we could exploit the linearity of $\Lambda(t)$ to observe that if $c_2(0)$ \emph{vanishes}, the troublesome term is removed entirely. So this state, at least, is certainly forward compatible: positing that the initial state is uncorrelated thus suffices to guarantee that the Boltzmann equation holds for $0<t \ll T$. However, it is much stronger than is required (we expect that a great many other choices of correlation are also forward compatible) and somewhat physically implausible (a totally uncorrelated dilute gas will swiftly build up some correlations, and indeed we can write an explicit equation for them; if they really were negligible at all times, the Boltzmann equation would reduce to the Vlasov equation). And again, this assumption fails to engage with the heuristic arguments for the vanishing of $\Lambda(t)\cdot c_2(0)$ for `reasonable' initial states.
\item An intermediate strategy (suggested in \citeN{wallacelogic}) is to assume \emph{Simplicity}: the assumption that the initial state's correlations can be described in a reasonably simple mathematical form. (We need to explicitly bar descriptions that involve starting with a simple sate, evolving it forward, and then time-reversing it!) We have heuristic, but extremely strong, grounds to assume that Simple states are forward compatible.
\end{enumerate}
The second and third of these assumptions, counter-intuitively, are time-reversal invariant: Simple states time reverse to simple states, uncorrelated states to uncorrelated ones. But they are not time \emph{translation} invariant: their forward \emph{and backward} time evolutes violate the condition. Imposing either condition guarantees (or, for the simplicity assumption, is heuristically likely to guarantee) that the Boltzmann equation holds for $0<t\ll T$, and that its time reverse holds for $0>t>-T$. So such conditions can only be imposed at the beginning of a system's existence, on pain of experimental disconfirmation.

Superficially, it is easy to make the argument that simplicity, at least, can naturally be expected from a physical process that creates a system: such processes cannot plausibly be expected to generate delicate patterns of correlations. (The same might be said for the straightforward assumption of forward compatibility.) But again, this is not innocent: the time reversal of a system's final state, just before its destruction, is certainly not forward compatible, and so our argument builds in time-directed assumptions. Assuming that we seek a dynamical explanation for irreversibility (and are not content, for instance, to appeal to unanalysable notions of agency or causation) then a familiar regress beckons, and we are led to assume some condition analogous to forward compatibility or simplicity, applied to the very early Universe. (I discuss this issue in more detail in \citeN{wallacelogic}.)

In any case, for the purposes of \emph{this} paper, what matters is that we have identified the time-asymmetric assumption being made in (a modern derivation of) the Boltzmann equation: it is an assumption about the initial condition of the system, phrased probabilistically (it is a constraint on the two-particle correlation function, which is inherently probabilistic) and with no implications as to the bulk distribution of particles in the system (insofar as these are coded by the one-particle marginal, given that all higher-order correlations are small). It will be instructive to compare this to the situation in Boltzmann's own derivation of the Boltzmann equation.

\section{The Boltzmann equation: contrast with the historical approach}\label{boltzmann-contrast}

The equation derived by Boltzmann himself\footnote{Here I follow Brown \emph{et al}'s~\citeyear{brownboltzmann}'s account of Boltzmann's work; see that paper for original references.} has the same functional form as what I have called the Boltzmann equation (albeit Boltzmann confined his attention to the case where $\rho_1$ is spatially constant, so that the free-particle term $\pb{H_1}{\rho_1}$  vanishes). But its interpretation is quite different. To Boltzmann, $\rho_1$ was not a probability distribution: it was a smoothed version of the \emph{actual} distribution function of particles over 1-particle phase space, so that $\rho_1(\vctr{q},\vctr{p})\delta V$ is the actual fraction of particles in a region $\delta V$ around $(\vctr{q},\vctr{p})$. (The smoothing is necessary because, with a finite number of point particles, $\rho_1$ would otherwise just be a sum of delta functions.)

Boltzmann makes a number of simplifying assumptions about the dynamics (notably: that collisions are hard-sphere events; that three-body collisions can be neglected; that long-distance interactions can be neglected) which are time-reversal invariant and broadly equivalent to the time-reversal-invariant assumptions made in the modern derivation. He then assumes the famous \emph{Stosszahlansatz} (SZA): the assumption that the subpopulation of particles which are about to undergo a collision has the same distribution of momenta as the population as a whole. Given this assumption, \emph{imposed over a finite time interval $[0,\tau]$}, he then deduces that over that time interval the Boltzmann equation holds. The assumption is thus explicitly time-reversal-noninvariant: assuming that $\rho_1$ is not stationary, if it holds over an interval then its time reverse cannot hold over that interval. (If it did, then both the Boltzmann equation and its time reverse would hold over that interval, contradicting Boltzmann's famous $H$-theorem, which tells us that the function $H[\rho]$ monotonically increases during the evolution of any non-stationary distribution under the Boltzmann equation.)

From a formal perspective, the probabilistic derivation in section \ref{boltzmann-modern} has significant advantages over the historical approach. (Perhaps unsurprisingly, given that it has in fact largely supplanted the historical approach in theoretical physics.) In particular:
\begin{enumerate}
\item The SZA is a condition that must hold \emph{throughout an interval of time} if that system is to obey the Boltzmann equation over that interval of time. But whether a condition holds at time $t>0$ is dynamically determined by the system's state at time $0$, and Boltzmann's derivation provides little insight as to what constraint on the state at time 0 suffices to impose the SZA at later times. Lanford's celebrated proof of the Boltzmann equation (Lanford~\citeyear{lanford1,lanford2,lanford2}; carefully discussed by \citeN{valentelanford} and \citeN{uffinkvalentelanford}) demonstrates that imposing (a generalisation of) the SZA at time 0 suffices to guarantee that it holds over some $[0,\tau]$ --- but $\tau$ is extremely short, only a fifth of a mean free collision time. 

By contrast, the probabilistic approach provides an explicit condition --- $\Lambda(t)\cdot c_2(0)\simeq 0$ for $0<t<T$ --- for when the Boltzmann equation holds, and allows us to state in closed form at least one correlation function --- $c_2(0)=0$ --- that guarantees that the condition will continue to hold. (To be fair, Lanford's results are at a much higher level of rigor than the probabilistic derivation, so a reader's assessment of this point will depend on their degree of tolerance of the mathematical practices of mainstream theoretical physics.)
\item The framework of the probabilistic approach readily generalises. On slightly different assumptions about the dynamics, for instance, it leads to Landau's kinetic equation (governing weak collision processes) or to the Lenard-Balescu equation (governing collisional plasmas). Boltzmann's original approach does not seem to have offered a comparably effective framework for the construction of other statistical-mechanical equations.
\item Most significantly, and as I will discuss further in section \ref{boltzmann-quantum}, the probabilistic approach readily transfers to quantum mechanics, where Boltzmann's original approach appears to fail completely.
\end{enumerate}

Against this, Boltzmann's original approach might be thought to have a major \emph{conceptual advantage} over the probabilistic approach, precisely because it eschews problematic notions of probability that seem to have no natural place in classical mechanics given that the latter features a deterministic dynamics. (Harvey Brown presses the point in his contribution to this volume.) To be blunt (the critic might say) then if progress in theoretical physics has moved from Boltzmann's original, relative-frequency, conception of the Boltzmann equation to a conception that relies on a mysterious concept of objective probability, then so much the worse for progress.

In my view, the ultimate resolution of this problem is quantum-mechanical: when we consider quantum statistical mechanics, these probabilities receive a natural interpretation. But even setting quantum theory aside, it is not obvious that the criticism is well-founded. For one thing, probabilistic notions are frequently appealed to in discussions of the Boltzmann equation, even where the latter is understood \emph{a la} Boltzmann: see Brown \emph{et al} \citeyear{brownboltzmann} as regards Boltzmann's own derivation, and \citeN{uffinkvalentelanford} in the context of Lanford's more rigorous analysis. So it is not after all clear that some notion of probability is not required --- in which case, why object to incorporating it into the equations themselves? 

More importantly, conceptual analyses have to answer to the actual shape of theoretical physics, at least insofar as the latter is empirically successful. I have already noted that several generalisations of the Boltzmann equation can be derived within the probabilistic framework but (so far as I know) have not been demonstrated within Boltzmann's conception of the equation. 

Now, these equations, whatever their derivation, are expressions of the one-particle marginal and can be reinterpreted without empirical consequence as equations in the actual relative frequency of particles' phase-space locations. But classical statistical mechanics \emph{also} offers examples where the predictions of the theory are themselves probabilistic in nature. In the next section, I present one such example; it will also serve as a demonstration that the analysis of irreversibility in section \ref{boltzmann-modern} is more general than just the Boltzmann equation.

\section{Mori-Zwanzig projection and the Fokker-Planck equation}\label{fokkerplanck}

Let's now consider a different Hamiltonian:
\be 
H(Q,P,q,p)= \frac{P^2}{2M}+V(Q) +\sum_{1\leq i \leq N}\omega_i(q^{i2}+p_i^2)+ \sum_{1\leq i \leq N}\lambda_i Qq^i
\ee
(where functional dependence on $q,p$ schematically depicts dependence on all of the $q^i,p_i$).
This describes a single particle in one dimension (with position $Q$ and momentum $P$) (the `system') interacting with a bath of harmonic oscillators (the `environment'); it is one common model for Brownian motion. Again we introduce probabilities explicitly via a probability density $\rho$ over the phase space for the $N+1$ particles, and take Liouville's equation as the basic dynamical equation for this system. That distribution can be decomposed as
\be 
\rho(Q,P,q,p)=\rho_S(Q,P)\rho_E(q,p) + C(Q,P,q,p)
\ee
where $\rho_S$ and $\rho_E$ are, respectively, the marginal probability distributions over system and environment, and $C$ is the correlation function between the two.

Heuristically we might hope to find:
\begin{itemize}
\item that there is an autonomous dynamics for the single particle (assuming that it is massive compared to the bath particles and that there are very many of the latter, so that their influence is a kind of background noise);
\item that the environment marginal $\rho_E$ is pretty much constant during the system's evolution, provided that it starts off invariant under the self-Hamiltonian 
\be
H_E = \sum_{1\leq i \leq N}\omega_i(q^{i2}+p_i^2)
\ee
of the bath. 
\end{itemize}
With that in mind, we define the following projection map on the space of distributions:
\be 
J \rho (Q,P,q,p) = \rho_S(Q,P) \mc{E}(q,p).
\ee
Here $\rho_S$ is the system marginal distribution of $\rho$, and $\mc{E}$ is some \emph{fixed} distribution for the environment, satisfying
\be 
\pb{H_E}{\mc{E}}=0.
\ee
(Typically we take $\mc{E}$ to be the canonical distribution for some given temperature.)
Then we can write $\rho$ itself as
\be 
\rho = J \rho + (1-J) \rho \equiv \rho_r + \rho_i.
\ee
This is actually a special case of a general process --- the \emph{Mori-Zwanzig projection}  --- for constructing autonomous dynamical equations in statistical mechanics. The components $\rho_r = J \rho$ and $\rho_i =(1-J)\rho$ are called, respectively, the \emph{relevant} and \emph{irrelevant} parts of $\rho$, and the hope is that an autonomous dynamics can be found for $\rho_r$ alone --- given certain assumptions about $\rho_i$. The decomposition of the two-particle marginal in the Boltzmann distribution into a product of one-particle marginals and a residual correlation term has a similar structure -- but in that case, the projection that defines the decomposition is nonlinear).

Without any approximation, we can obtain an integro-differential equation for $\rho_r$ somewhat similar in form to equation (\ref{protoboltzmann}), obtained in our analysis of the Boltzmann equation:
\begin{eqnarray}
\ddt\rho_r(t)&=&J L_H \rho_r(t) \nonumber
\\
&+ &J \int_0^t \dr{\tau}  \e{(1-J)\tau L_H}(1-J)L_H \rho_r(\tau) \nonumber
\\
&+& J L_H  \e{(1-J)t L_H} \rho_i(0),
\end{eqnarray}
where
\be 
L_H \rho \equiv \pb{H}{\rho}.
\ee
This equation is exact, and so displays no irreversibility. However, if $J$ has been chosen appropriately then we would hope to find:
\begin{itemize}
\item That the kernel in the second term falls off sufficiently rapidly with $\tau$ that it can be approximated as
\begin{eqnarray}
\int_0^t \dr{\tau}  \e{(1-J)\tau L_H}(1-J)L_H \rho_r(\tau) & \simeq & \left( \int_0^T \dr{\tau}  \e{(1-J)\tau L_H}(1-J)L_H\right)\rho_r(t) \nonumber \\
&\equiv& K \rho(t)
\end{eqnarray}
where $T$ is some very long time of order the recurrence time.
\item That the third term can plausibly be expected to vanish given some `reasonable' constraints on $\rho_i(0)$.
\end{itemize}
Given the first assumption, and defining 
\be
\Lambda(t)=J L_H  \e{(1-J)t L_H},
\ee
we are left with the equation
\be 
\ddt\rho_r(t)=J L_H \rho_r(t)+ K \rho_r(t) + \Lambda(t) \rho_i(0)
\ee
which is very similar in form to the proto-Boltzmann equation (\ref{protoboltzmann2}). No time-asymmetric assumption has been made to get this far; we now have to make an initial-time assumption about $\rho_i(0)$ to guarantee that $\rho_i(0)$ is forward compatible, \iec that $\Lambda(t)\rho_i(0)\simeq 0$ for $0<t<T$. As with the Boltzmann case, we can guarantee this by taking $\rho_i(0)=0$; as with the Boltzmann case, this is usually overkill.

Returning to the specific case of the oscillator bath, we can calculate the second term approximately by working to second order in perturbation theory. Given a forward-compatible initial state (which, in this case, will typically require at least that $\rho_E(0)=\mc{E}$), we get the equation
\begin{equation}\label{FP}
\ddt \rho_S(t) = \{\tilde{H}_S,\rho_S(t)\}\nonumber +  \eta \pb{X}{P\rho_S(t)}+ \alpha \pb{X}{\pb{X}{\rho_S(t)}} + f \pb{X}{\pb{P}{\rho}}
\end{equation}
where 
\be
\tilde{H}_S = P^2/2M + V(Q)+ \Delta V(Q)
\ee
and $\xi, \eta$, $f$ and $\Delta Q$  are (somewhat complicated) functions of the $\lambda_i$ coefficients and the frequencies $\omega_i$ of the oscillator bath. For reasonable assumptions, $f$ is usually negligible; equation (\ref{FP}) is then the \emph{Fokker-Planck equation}, and we can get insight into it by recognising that it is the equation for the probability distribution of a particle evolving under the \emph{Langevin equation}, the stochastic differential equation
\be 
\dot{Q(t)}=P(t)/M; \,\,\,\, \dot{P}(t)=- \eta P(t) + \xi(t)
\ee
where $\xi(t)$ is a random variable satisfying
\be 
\langle \xi(t)\rangle =0; \langle \xi(t_1)\xi(t_2)\delta t^2\rangle =2 \alpha \delta t.
\ee
(For details of both, see \citeN[chs.1-2]{zwanzigbook}.)

These equations --- which are highly effective at describing the physics of Brownian motion --- make explicitly probabilistic predictions: for instance, that the root-mean-square value of the particle's distance from its starting place after time $t$ is (after an initial transient phase) equal to $2\alpha t/\eta^2$. Unlike the case of Brownian motion, the equations cannot be reinterpreted as non-probabilistic equations concerning the relative frequency of particles in a single system.

Incidentally, since the underlying Hamiltonian of this system is quadratic, the appeal to second-order perturbation theory is dispensible: with a bit of care, we can solve it \emph{exactly} and confirm the approximate validity of equation (\ref{FP}). See \citeN{pazzurekreview} for details.

\section{Classical statistical mechanics: could do better?}\label{coulddobetter}

I began with the observation that non-equilibrium statistical mechanics is concerned with establishing the relations between dynamical equations at different levels of description, especially in the cases where the higher-level equations are probabilistic and/or irreversible; I noted that it is mysterious how probability or irreversibility can be derived from an underlying dynamics which has neither feature, but that paying attention to the details of such derivations might be illuminating.

We have now seen two such derivations in detail (conceptual detail at any rate; the mathematics was left schematic), and I have argued that
\begin{enumerate}
\item The probabilistic features of the equations of statistical mechanics are not readily removable: even in the case of the Boltzmann equation, which can be reinterpreted as a non-probabilistic equation, its derivation is better understood and more readily generalised when understood probabilistically, and in the case of the Fokker-Planck and Langevin equations, the equations cannot even be reinterpreted non-probabilistically.
\item The irreversibility of at least a wide class of classical statistical-mechanical equations (specifically, those which are either derived from the BBGKY hierarchy via the same approximation scheme we used for the Boltzmann equation, or derived by the Mori-Zwanzig method from a linear projection) can be tracked to a cleanly-stated assumption about the initial microstate --- an assumption, however, which is again stated probabilistically.
\item No clue has been gleaned about the origins of the probabilities used in these equations: in each case, our starting point was the Louville equation, applied to a probability distribution placed by \emph{fiat} on phase space.
\end{enumerate}
So we are in the unsatisfactory position of having acquired considerable evidence as to the importance and ineliminability of probabilities in non-equilibrium statistical mechanics, without gaining any insight at all into the origins of these probabilities. In the rest of the paper, I will demonstrate that this dilemma is resolved, or at least radically transformed, when we move from classical to quantum mechanics.

\section{Mixed states and probability distributions in quantum statistical mechanics}\label{mixedstates}

At first sight, there is a straightforward translation scheme between classical and quantum that maps the previous section's results directly across to quantum theory. To phase-space points correspond Hilbert-space rays. To Hamilton's equation corresponds Schr\"{o}dinger's. To probability distributions on phase space correspond density operators. To Liouville's equation corresponds its quantum counterpart,
\be \label{quantum-liouville}
\dot{\rho}=L_H \rho \edf \frac{i}{\hbar}\comm{\rho}{H}.
\ee
Appearances, however, can be deceptive. 

Von Neumann originally \emph{introduced} the density operator to represent ignorance of which quantum state a system is prepared in: the idea is that if the system is prepared in state $\ket{\psi_i}$ with probability $p_i$, then for any measured observable $\op{X}$ its expectation value is
\be 
\langle \op{X}\rangle=\sum_i p_i \matel{\psi_i}{X}{\psi_i}=\tr\left[\op{X}\left(\sum_i p_i \proj{\psi_i}{\psi_i}\right)\right],
\ee
so that if we define 
\be \label{density-operator}
\rho=\sum_i p_i \proj{\psi_i}{\psi_i},
\ee
the formula $\langle\op{X}\rangle=\tr(\op{X}\rho)$ neatly summarises our empirical predictions. But it was recognised from the start that (\ref{density-operator}) cannot be inverted to recover the $p_i$ and $\ket{\psi_i}$: many assignments of probabilities to pure states give rise to the same $\rho$. By contrast, for any two distinct probability distributions over phase space there is (trivially) some measurement whose expectation value is different on the two distributions.

(It is true that if we add the extra information that the $\ket{\psi_i}$ are orthogonal, and if we assume that $\rho$ is non-degenerate, then the $p_i$ and the $\ket{\psi_i}$ can after all be recovered. But (i) there is no particular problem in making sense of probability distributions over non-orthogonal states; (ii) the density operators used in statistical mechanics --- particularly the microcanonical and canonical distributions --- are massively degenerate.)

That's a puzzling disanalogy, to be sure, but hardly fatal. In these post-positivist days, surely we can cope with a little underdetermination? But it gets worse. For recall: a joint state of system $A$ and system $B$ is generically \emph{entangled}: it cannot be written as a product of pure states, and indeed there is no pure state of system $A$ that correctly predicts the results of measurements made on system $A$ alone. Indeed, the correct mathematical object to represent the state of system $A$ alone, when it is entangled with another system, is again a density operator $\rho$, obtained by carrying out the partial trace over the degrees of freedom of system $B$. 
(But \emph{this} density operator cannot be understood as a probabilistic mixture of pure states, on pain of failing to reproduce in-principle-measureable Bell-type results.)

The corollary is that we can only represent a quantum system (or, indeed, our information about a quantum system, if that's your preferred way of understanding probabilities here) by a probability measure over pure states if with probability 1 the system is not entangled with its surroundings. And in the case of the macroscopically large systems studied in statistical mechanics, this is a wildly implausible assumption: even if for some (already wildly implausible) reason we were confident that at time 0 there was no entanglement between system and environment, some will develop extremely rapidly. (But note, conversely, that simply assuming (or idealising) that the system is currently dynamically isolated from its surroundings in no way rules out the possibility that it is entangled with them.)

If we want to place a probability measure over states of a system, then, in general the only option will be to place it over the system's possible \emph{mixed} states. Mathematically, that's straightforward enough: if a system has probability $p_i$ of having mixed state $\rho_i$, then the expected value of a measurement of $\op{X}$ is
\be 
\langle \op{X}\rangle = \sum_i p_i \tr(\op{X}\rho_i)=\tr\left[\op{X}\left(\sum_i p_i \rho_i\right)\right].
\ee
But is clear that we get the \emph{exact same} predictions by simply assigning the system the single mixed state $\sum_i p_i \rho_i$. In other words, probability distributions \emph{over} mixed states are indistinguishable from \emph{individual} mixed states. 

Note how different this is from the classical case. Even formally (never mind conceptually), the move from phase-space dynamics to distributional dynamics is an extension of the mathematical framework of classical physics. But in quantum mechanics, the move from pure states to density operators is forced upon us by considerations of entanglement quite independently of any explicit probabilistic assumption, and so there is nothing formally novel about their introduction in statistical mechanics.

Put another way, \emph{any mathematical claim} in quantum statistical mechanics can be interpreted as a claim about the actual (perhaps mixed) state of the system. Nothing formally requires us to interpret it as any kind of probability distribution over quantum states.

Now, perhaps this would be unimpressive if the foundations of quantum statistical mechanics made major \emph{conceptual} use of probabilistic ideas in deriving their results: if, for instance, the arguments given for the mathematical form of quantum statistical equilibrium relied on probabilistic ideas. But a casual perusal of the (fairly minimal) literature on quantum statistical mechanics serves to disabuse one of this notion. From Tolman's classic text~\citeN{tolman} to modern textbook discussions, the norm is to start with classical statistical mechanics  and then move from a probability distribution over phase-space points to a probability distribution over \emph{eigenstates of energy drawn from a particular basis}. I don't really know any way to make sense of this, but \emph{certainly} it can't be made sense of as a probability distribution over possible states of the system given that (a) there is absolutely no reason, given the massive degeneracy that typically characterises the Hamiltonian of macroscopic systems, to prefer one energy basis over another; (b) there is no a priori reason, even ignoring entanglement, to assume with probability 1 that the system is in an eigenstate of energy; (iii) we shouldn't ignore entanglement anyway, so assuming the system is in \emph{any} pure state is unmotivated.\footnote{Very occasionally (the only place I know is Binney \emph{et al}~\citeyear{binney}) it is observed that a probability distribution uniform over all pure states with respect to the natural(unitary-invariant) measure on states gives the same density operator as a distribution uniform over a particular  orthonormal basis, and that this is the justification for the latter; even here, though, the problem of mixed states is left unanswered.}

This is not to say that there is not high-quality, rigorous work on the foundations of quantum statistical mechanics, or even on the form of the equilibrium distributions (for recent examples see Goldstein \emph{et al}~\citeyear{goldsteinequilibrium}, Malabarba \emph{et al}~\citeyear{malabarba}, and \citeN{shortequilibrium}). But insofar as these arguments avoid the above fallacies, they do so because they try to establish (or at any rate can be interpreted as trying to establish) that the actual mixed state of a system at equilibrium is the microcanonical or canonical ensemble, not that the correct probability distribution over such states reproduces that ensemble.

I conclude that there is no justification, and no need, for interpreting the mixed states used routinely in quantum statistical mechanics as any kind of probability distribution over quantum states, rather than as the (mixed) quantum states of individual systems. And if that is the case, then --- since in the real world ``classical'' systems are just quantum systems that we can get away with approximating as classical ---  it becomes at least very tempting to interpret the `probability' distributions of classical mechanics simply as classical limits of individual quantum states, without any need at all for a distinctively statistical-mechanical conception of probability.

In fact --- at least for the non-equilibrium systems we have considered so far --- this move is not just tempting, but compulsory. As we will see, it is simply not viable to interpret these systems as probability distributions over localised quasi-classical systems, once quantum mechanics is taken into account.

\section{The quantum mechanics of dilute gases}\label{dilutegassection}

The dilute gas of hard spheres is the paradigm example in classical statistical mechanics, both in considering the approach \emph{to} equilibrium (the Boltzmann equation assumes this system) and in analysing equilibrium itself. But what happens when quantum mechanics is used to analyse a system of this kind? 

Let us begin with some parameters. The typical gas at standard temperature and pressure has a molecular mass of around $10^{-26}$ or $10^{-27}$ kg, a molecular scale of around $10^{-10}$ m, a density of about 1 $\mathrm{kg}\,\mathrm{m}^{-3}$, and typical molecular velocities of around $10^2$ or $10^3$ $\mathrm{m}\,\mathrm{s}^{-1}$; for definiteness, let's assume our gas is confined in a 1$\mathrm{m}^3$ box, and that it consists of $10^{27}$ hard spheres of mass $10^{-27}$ kg and cross-sectional area $10^{-20} \mathrm{m}^2$ moving at a mean speed of $10^{3}$ $\mathrm{m}\,\mathrm{s}^{-1}$.

If classical \emph{micro}dynamics is to be a good approximation for this system, presumably this means that the particles in the gas must be, and remain, localised. So let's start the system off in some (probably unknown) state in which each molecule is a localised wavepacket with a reasonably definite position and momentum. We can now use approximately classical means to consider how those wavepackets evolve prior to their first collisions.

Elementary methods tell us that for a gas with these parameters, the mean free path --- the mean distance travelled by a given particle before it collides with another --- is $10^{-7}$m. The typical particle will cover this distance in $10^{-10}$ seconds.

How wide will its wavepacket be at that time? Let its original width be $\Delta q(0)$, so that its initial spread $\Delta p$ in momentum is at least $\sim \hbar/\Delta q(0)$. Roughly speaking, its additional spread in position as a result of this spread in momentum after time $t$ will be $t \times \Delta p/m$, so that its total spread after time $t$ is given by
\be 
\Delta q(t)=\Delta q(0)+t \Delta p/m = \Delta q(0)+\frac{\hbar t}{m} \frac{1}{\Delta q(0)}
\ee
Elementary calculus then tells us that this is minimised for $\Delta q(0)=\sqrt{\hbar t/m}$, so that 
\be 
\Delta q(t)\geq 2 \sqrt{\hbar t/m}.
\ee
For the parameters governing our ideal gas, this means that the minimum size of the packet after collision is approximately $3\times 10^{-8}$ m.\footnote{Don't be fooled into thinking that `decoherence' will somehow preserve localisation to a greater degree than this. These are microscopic particles: they don't decohere, or rather: their decoherence is caused by their collision with other particles, and we're explicitly considering the between-collision phase.}

But this is thirty times the diameter of the molecule! To a rough approximation, the first set of scattering events in our dilute gas will be well modelled by \emph{plane-wave} scattering off a hard-sphere scattering surface. The resultant joint state of two scattering particles will be a sum of two terms: one (much the larger one) corresponding to the state of the two particles in the absence of their interaction, and one an entangled superposition of the particles after scattering, with significant amplitude for any direction of scattering. Each particle's individual mixed state will be highly delocalised.

This is fairly obviously nothing like the classical microdynamics, in which each pair of particles deterministically scatters into a specific post-collision state determined entirely by the pre-collision state. A quantum-mechanical dilute gas, even if it is initially highly localised, rapidly evolves into a massively entangled and massively delocalised mess. So the classical microdynamics that underpins the derivation of Boltzmann's equation (either in the `modern' form I presented in section \ref{boltzmann-modern}, or in Boltzmann's original derivation) is \emph{wildly false} for real, physical gases. 

Let's be clear about what we have learned here. The point is not that there are quantum regimes where the classical Boltzmann equation breaks down. There are such regimes, of course: the statistical mechanics literature is replete with discussions of ``quantum'' gases, but in general this refers to situations where the intermolecular interaction has some complex quantum form, or (more usually) where the gas is sufficiently dense that the effects of Bose-Einstein or Fermi-Dirac statistics come into play, or where (as in the photon or phonon gas) particle number is not conserved. But I am not considering that situation; nor am I considering the situations where the \emph{quantum} Boltzmann equation is appropriate (as discussed by Brown, this volume). There is abundant empirical evidence that the particle distribution in ordinary, nonrelativistic, dilute gases is governed by the \emph{classical} Boltzmann equation --- and yet, we have seen that in those gases, the microdynamics 
assumed in the derivations of that equation is simply incorrect.

Why, then, do these derivations work? To address this question, it will be useful to adopt a representation of quantum-mechanical systems as functions on phase space (as distinct from the usual representations on configuration or momentum space); doing so will also cast light more generally on the relation between quantum states and classical probability distributions.

\section{The quantum/classical transition on phase space}\label{wignersection}

In the position representation of a quantum state, information about the probability of a given result on a \emph{position} measurement can be readily read off the quantum state via the mod-squared amplitude rule, whereas information about results of \emph{momentum} measurements are encoded rather inaccessibly in the phase variation of the wavefunction. In the momentum representation, the reverse is true. A phase space representation can be seen as enabling us to more readily access both lots of information.

An alternative way of understanding quantum mechanics on phase space is more direct: if the position representation is optimised for position measurements, and the momentum representation for momentum measurements, how can we represent the state in a way optimised for phase-space measurements? One immediate objection is that the uncertainty principle forbids joint measurements of position and momentum; in fact, though, it forbids only \emph{sharp} joint measurements. The by-now well-established formalism of ``positive operator valued measurements'' (POVMs) allows for a wide variety of unsharp measurements, including phase-space measurements. 

Indeed, we can straightforwardly write down such a family of phase space measurements: choose any wave-packet state $\ket{\Omega}$, reasonably well localized in position and momentum but otherwise arbitrary (Gaussians are good choices) and let $\ket{\vctr{q},\vctr{p}}$ denote the state obtained by translating the original state first by $\vctr{q}$ in position and then by $\vctr{p}$ in momentum. (Here $\vctr{q}$ and $\vctr{p}$ are vectors representing the $M$ position and $M$ momentum coordinates of our system; if it consists of $N$ particles, for instance, $M=3N$.) Then the (continuously infinite) family of operators $(2 \pi)^{-M}\proj{\vctr{q},\vctr{p}}{\vctr{q},\vctr{p}}$ is a POVM. (See appendix for proof; the equivalent result for Gaussian choices of the wavepacket is standard, but I am not aware of a proof in quite this form.)
Since these wavepacket states are not orthogonal, they do not define a \emph{sharp} measurement; however, if we partition phase space into disjoint cells $C_i$ which are large compared to the spread of the wave-packet state (and thus, at a minimum, large compared to $\hbar^M$), the operators
\be 
\op{\Pi}_i=\int_{C_i}\proj{\vctr{q},\vctr{p}}{\vctr{q},\vctr{p}}
\ee
will approximately satisfy
\be 
\op{\Pi}_i\op{\Pi}_j \simeq \delta_{ij}\op{\Pi}_i.
\ee
The \emph{Husimi function}, defined for quantum state $\rho$ by
\be 
H_\rho(\vctr{q},\vctr{p})=(2\pi)^{-M}\bra{\vctr{q},\vctr{p}}\rho\ket{\vctr{q},\vctr{p}},
\ee
can thus be consistently interpreted as a phase-space probability distribution  provided we do not probe it on too-small length scales. In fact, the Husimi function can be inverted to recover $\ket{\psi}$, so in principle we could use it as our phase-space representation; however, for our purposes it is inconvenient, and a better choice is the \emph{Wigner function},
\be 
W_\rho(\vctr{q},\vctr{p})=\frac{1}{(\hbar \pi)^M}\int \dr{\vctr{x}}\e{2i\vctr{p}\cdot \vctr{x}/\hbar}\bra{\vctr{q}\!-\!\vctr{x}}\rho\ket{\vctr{q}\!+\!\vctr{x}}.
\ee
The mathematical form of the Wigner function is not transparently connected with anything physical, but in fact, if we use the Wigner function (which is real, though not necessarily positive) as a probability measure, it will approximately give the correct phase-space measurement probabilities if averaged over regions large compared to $\hbar^M$. (Indeed, the Husimi function can be obtained from the Wigner function merely by smearing it in position and momentum with the respective mod-squared wavefunctions of the wavepacket state; one advantage of the Wigner function is that it abstracts away from the need to specify a particular choice of wavepacket state.)

The main virtue of the Wigner function is that its dynamics can be conveniently compared to the classical case. Recall that classical probability distributions evolve by Liouville's equation:
\be 
\dot{\rho}=\pb{H}{\rho}.
\ee
Transforming the Schr\"{o}dinger equation to the Wigner representation tells us that the Wigner function satisfies
\be 
\dot{W}=\mb{H}{W}\edf\frac{2i}{\hbar}\sin \left(\frac{\hbar}{2i}\pb{\cdot}{\cdot}\right)\cdot (H,W).
\ee
Here $\mb{\cdot}{\cdot}$, the \emph{Moyal bracket}, is best understood via its series expansion: assuming $H$ has the standard kinetic-energy-plus-potential-energy form, we can expand it as
\be 
\dot{W}=\mb{H}{W}=\pb{H}{W}+\frac{\hbar^2}{24}\frac{\partial^3V}{\partial \vctr{q}^3}
\frac{\partial^3W}{\partial \vctr{p}^3}+ O(\hbar^4).
\ee
So the dynamics of the Wigner function is the dynamics of the phase-space probability distribution, together with correction terms in successively higher powers of $\hbar$, which suggests that \emph{ceteris paribus}\footnote{The full story is more complicated, as stressed by \citeN{pazzurekreview}; see also my discussion in \citeN[ch.3]{wallacebook}.} the corrections become negligible for macroscopically large systems --- and are exactly zero for free particles in any case.

Let's pause and consider these technical results from a conceptual viewpoint. The natural question one ends up asking first, when confronted with something like the Wigner function, is: why can't we just suppose that this \emph{is} a classical probability distribution, and thus resolve the paradoxes of quantum theory? And the usual answer given is: the Wigner function is generally\footnote{Specifically, it's nonnegative for pure states only if those states are Gaussian; cf \citeN{hudson1974}.} not nonnegative, which  probability distributions have to be. But this is not the real problem (if it were, we could shift to the Husimi distribution, which \emph{is} reliably nonnegative). The real problem is that there is no underlying microdynamics on phase-space points such that the Wigner function's (or the Husimi function's) dynamics can be seen as a probabilistic dynamics for that microdynamics. 

Put another way, suppose we were simply given the Liouville equation. Our interpretation of that equation as the evolution equation for a probability distribution rests on the fact that we can get back the Liouville equation as the dynamics for probability distributions induced by Hamilton's equations. (That the latter are deterministic is not relevant; similarly, given the Fokker-Planck equation we can justify reading it probabilistically by observing that it is the probabilistic dynamics induced by the stochastic Langevin equation.)\footnote{A technical aside: at least formally, any linear map of a vector space to itself that preserves the $L^1$ norm and which preserves the subset of vectors with nonnegative coefficients in a given basis can be interpreted as generated from a possibly-stochastic dynamics with respect to this basis. So what is doing the technical work here is the fact that neither the Wigner function nor the Husimi function have dynamical equations which can be extended to the space of all functions on phase space without violating the basis-preservation rule, \iec without mapping some positive functions to negative ones. In the case of the Wigner function, the dynamical rule fails to preserve positivity even on the subspace of functions that correspond to quantum states, but this is not a requirement.}

The point is this: the Wigner function is not a function on phase space in the \emph{original} understanding of phase space. It is not a function on a space which can be \emph{physically interpreted} as the space of possible positions and momenta of point particles, for there are no such particles. It does, however, illustrate that in certain regimes (those where the higher-order terms in its evolution can be neglected, and in which the system is not probed on length-scales where the uncertainty principle comes into play), there is an approximate isomorphism between the dynamics of the quantum \emph{state} and that of the classical \emph{probability distribution}. The classical limit of quantum mechanics, to quote \citeN{ballentine}, is classical \emph{statistical} mechanics. 

To strengthen the point, consider the Wigner (or Husimi) functions of pure states. They certainly do not correspond to delta functions on phase space, but to extended distributions on phase space. Even quantum wave-packet states are taken to Gaussian packets, not to delta functions: to probability distributions (or quantum generalisations thereof), not to individual microstates. 

Now: in certain contexts, those wavepackets avoid spreading out and evolve so as to mirror the evolution of phase-space \emph{points} under classical microdynamics. In \emph{these} contexts, it is perhaps reasonable to think of classical \emph{microdynamics} as emerging as a limiting regime of quantum physics. But the contexts in which this occurs are relatively narrow (corresponding to isolated systems with large masses and non-chaotic dynamics; cf \citeN{zurekpazchaos} and \citeN[ch.3]{wallacebook}) and do not include many of the typical contexts in which ``classical'' statistical mechanics is applied, as we saw in the previous section.

\section{The classical Boltzmann equation from a quantum perspective}\label{boltzmann-quantum}

In the limiting case of a non-interacting dilute gas of atoms, Hamiltonian microdynamics is wildly false --- yet Liouville dynamics is exactly correct. This suggests a route to understanding the classical Boltzmann equation in a quantum universe: if Liouvillian dynamics (applied, of course, to Wigner functions) is \emph{approximately correct} even for a dilute interacting gas, then the modern derivation of the Boltzmann equation via the BBGKY hierarchy will still go through. And it is heuristically plausible that Liouvillian dynamics are approximately correct for such systems: after all, the particles in the gas spend most of their time moving freely, and in the free-particle regime Liouvillian and quantum dynamics coincide.

In fact, we can easily go beyond heuristic plausibility. The BBGKY hierarchy, and the various marginals, can be defined for the Moyal-bracket dynamics and the N-particle Wigner function \emph{exactly} as for the Poisson-bracket dynamics and the classical phase-space distribution: the functional form is identical in both cases. As such, we can replicate the derivation of the Boltzmann equation by replacing Poisson with Moyal brackets throughout. Making the same assumptions and approximations at each step, we again obtain equation (\ref{protoboltzmann}):
\[ 
\ddt{\rho_1}(t) = \{H_1,\rho_1(t)\}  + \kappa[\sigma,\rho_1(t)] + \Lambda(t) \cdot c_2(0),
\]
which reduces to the Boltzmann equation if the forward-compatibility condition $\Lambda(t) \cdot c_2(0)=0$ is satisfied.
 
\emph{Formally} speaking, this differs from the classical Boltzmann equation only in that $\sigma$ is the quantum-mechanical, and not the classical, cross-section. But in fact this is exactly what we need to explain the applicability of the Boltzmann equation: in actual usage, the cross-section used is invariably that calculated by quantum mechanics if quantum and classical predictions differ. (In the physics of nuclear or chemical reactions, for instance, we use a generalisation of the Boltzmann equation to handle multiple particle species, and include quantum-derived cross sections for reactions that turn particles of one species into particles of another.)

Conceptually, this equation is completely different, though: $\rho_1$, and $c_2$, are features not of a probability distribution over an unknown classical microstate, but of a single quantum state. There is not even any particular requirement for that state to be a pure state (albeit that the `overkill' condition, $c_2=0$, does require the true $N$-particle state to be mixed). In particular, there is nothing probabilistic about the forward-compatibility requirement: it is a constraint on the actual quantum state. 

Or, better: there \emph{is} something probabilistic about $\rho$, and about the forward-compatibility requirement, but only in the sense that there is something probabilistic about the quantum state itself (however that probabilistic nature is to be understood). The probabilities in the Boltzmann equation are not a new species of probability: they are Born-rule quantum probabilities, in the classical limit where the Liouville equation (though not Hamilton's equations) are approximately valid.

At this point, let us revisit section \ref{boltzmann-contrast}'s comparision between a BBGKY derivation of the Boltzmann equation and Boltzmann's own, probability-free, derivation. There, I suggested that the BBGKY derivation has the major virtues of (1) providing a single and relatively tractable initial-time condition for the Boltzmann equation to hold at later times; (2) generalising much more readily to other statistical-mechanical contexts, and (3) most importantly, applying to quantum mechanics; against this, I conceded that in the classical context it has a greater reliance on a somewhat mysterious notion of statistical-mechanical probability (albeit that notion is clearly necessary in classical statistical mechanics, as the Langevin equation also demonstrates).

From a quantum perspective, this disadvantage is entirely negated and the force of (3) becomes clear. In a quantum universe, the statistical-mechanical probabilities of the (classical) Boltzmann equation are nothing over and above quantum probability, something which (however mysterious it might be) we are already committed to. And in a quantum universe, the BBGKY derivation remains largely unchanged, while Boltzmann's original derivation relies on assumptions about the microdynamics of the system in question that are clearly, demonstrably, false.

\section{The Fokker-Planck equation from a quantum perspective}\label{fokkerplanck-quantum}

The Boltzmann equation is in practice used simply to make predictions about the relative frequency of the particle distribution, so that its probabilistic nature is somewhat obscured; I introduced the Langevin and Fokker-Planck equations in section \ref{fokkerplanck} to demonstrate that often the predictions (and not just the methods of derivation) of classical statistical mechanics are probabilistic: often the equations that emerge in non-equilibrium statistical mechanics are stochastic. In fact --- at least in this particular example --- those probabilities also turn out to be quantum-mechanical probabilities, so that (for instance) Brownian motion, in our universe, needs to be understood as a quantum-mechanically random process, not simply as a consequence of our ignorance of the classical microstate or something similar.

To elaborate: in section \ref{fokkerplanck} I derived the Fokker-Planck equation starting from the Liouville equation appropriate to a Hamiltonian that couples one large, with many small, harmonic oscillators, as a special case of the Mori-Zwanzig method of projection. The latter translates across to quantum mechanics \emph{mutatis mutandis}, simply be reinterpreting $\rho$ as a density operator and redefining
\be 
L_H \rho = - \frac{i}{\hbar}[H,\rho].
\ee
In the particular case of the coupled oscillators, since the system is quadratic its phase-space dynamics are \emph{identical} whether understood classically or quantum-mechanically, so that equation (\ref{FP}) can be taken directly over to the quantum context simply by replacing Poisson brackets with commutators and probability distributions with mixed states (and taking a little care with non-commuting operators):
\begin{eqnarray}
\ddt \rho_S(t) & = & -\frac{i}{\hbar} [\tilde{H}_S,\rho_S(t)]\nonumber +  \gamma \frac{i}{\hbar}[X,(P\rho_S(t)+ \rho_S(t)P)] \nonumber \\ & & - D [X,[X,\rho_S(t)]] - \frac{1}{\hbar}f[X,[P,\rho_S(t)]].
\end{eqnarray}
I have absorbed some factors of $\hbar$ into the coefficients, following the conventions of \citeN{pazzurekreview}. 

This is an equation for the reduced state $\rho_S(t)$ of the large oscillator, derived by separating the combined state $\rho(t)$ of large and small oscillators into
\be 
\rho(t) = \rho_S(t)\otimes \mc{E} + \rho_I(t),
\ee
for some fixed state $\mc{E}$ of the small oscillators, invariant under their self-Hamiltonian.
It is valid (to second order in perturbation theory, and after some short `transient' period for the coefficients to settle down to their long-term values) whenever $\rho_I(0)$ satisfies the forward compatibility conditions, and in particular whenever $\rho_I(0)=0$, \iec when the system and environment are initially uncorrelated. Quantum-mechanical features of the equation show themselves only when it is applied to highly non-classical \emph{states}, such as coherent superpositions of wavepackets in different regions: in that context, it is the standard \emph{decoherence master equation} of environment-induced decoherence, and rapidly decoheres that superposition. But given as input a wavepacket, or an incoherent mixture of wavepackets, then it exactly reproduces the classical probability dynamics.

Classically speaking (and assuming that the $f$ coefficient is negligible) we have already noted that those dynamics can be interpreted as the probability-distribution evolution defined by the stochastic Langevin equation. But in the classical case the probabilities arise entirely because the input environment state $\epsilon$ is a probability distribution: the apparently-stochastic behaviour of the system is a consequence of an explicitly probabilistic assumption about the environment. (Normally we assume $\mc{E}$ is the canonical distribution of the oscillators.) 

This is not the case in quantum mechanics. Firstly, even if we do choose a canonical state
\be 
\mc{E}_C= \frac{\e{-\beta H_E}}{\tr \e{-\beta H_E}}
\ee
for the environment, it is (as I argued in section \ref{density-operator}) debatable at best whether this state should be interpreted probabilistically. But suppose it is so interpreted: then we should actually regard $\mc{E}_C$ as a probabilistic mixture of energy eigenstates, and standard statistical arguments tell us that this probability distribution is dominated by eigenstates $\ket{\{n_i\}}$ (where $n_i$ is the excitation number of the $i$th oscillator) satisfying
\be 
\sum_i ( n_i + 1/2) \omega_i \simeq \langle H_E\rangle_{\mc{E}_C}.
\ee
Each such eigenstate is a perfectly valid choice for the environment state $\mc{E}$, and if we explicitly check the formulae for the coefficients in \citeN{pazzurekreview}, we find that for the overwhelming majority of such states, the coefficients have pretty much the same value as they would if calculated with the mixed state $\mc{E}_C$. 

But the initial state $\proj{\psi}{\psi}\otimes \proj{\{n_i\}}{\{n_i\}}$ is \emph{pure}, and (whatever the right interpretation of quantum-mechanical \emph{mixed} states) contains no purely statistical-mechanical probability assumptions. So the probabilities in the Fokker-Planck equation derived from that initial state are purely quantum-mechanical. And if we hold on (implausibly) to the statistical interpretation of $\mc{E}_C$, we find that its statistical-mechanical probabilities make virtually no contribution to the probabilities in \emph{its} Fokker-Planck equation. This is radically unlike the classical case, in which perfect knowledge of the environment's microstate suffices (in principle) to eliminate the probabilities entirely from the large oscillator's dynamics.

As a final observation: it is fairly easy to show that the equilibrium state of a Fokker-Planck oscillator is a thermal state at the same temperature as its bath. Since this result holds even when the environment's initial state is a pure state as above, at least in \emph{this} case the equilibrium state again has to be interpreted purely quantum-mechanically, with no trace of statistical-mechanical probabilities.

\section{Conclusion}\label{conclusion}

Classical non-equilibrium statistical mechanics, in its contemporary formulations and across at least the wide class of applications considered in sections \ref{bbgky}--\ref{fokkerplanck}, adds to Hamiltonian classical mechanics two additional components: probability (via a move from Hamiltonian to Liouvillian dynamics) and irreversibility.

Irreversibility is introduced by some kind of boundary condition; that much could have been deduced \emph{a priori} given that it is not present in the underlying dynamics (Hamiltonian or Liouvillian), but the common form of that condition in BBGKY-derived or Mori-Zwanzig-derived statistical mechanics differs from the two most-commonly-discussed such conditions in the philosophical literature:
\begin{itemize}
\item Unlike the commonly-discussed \emph{past hypothesis} that the initial state of the system (most commonly, the entire Universe) is low in entropy, the forward-compatibility conditions are conditions on the \emph{microstate}, invisible at the coarse-grained level. (Low entropy, insofar as that is a condition on the \emph{coarse-grained} state, is neither necessary nor sufficient for the Boltzmann or Fokker-Planck equations to hold.)
\item Forward compatibility is closer to the \emph{Stosszahlansatz} of Boltzmann's own approach to non-equilibrium statistical mechanics, which is also a condition on the microstructure of the system, invisible at the coarse-grained level. But forward compatibility is (i) a condition on the Liouvillian, not the Hamiltonian, state (and thus explicitly probabilistic); (ii) mathematically formulated as a condition on the initial state, not as a requirement that must hold throughout the evolution period; (iii) demonstrably satisfied by at least one class of initial states, namely those for which the `irrelevant' part of the state vanishes.
\end{itemize} 
Probability might be the most mysterious part of classical statistical mechanics as long as it is understood in isolation from quantum theory --- hence the temptation of Brown and others to eliminate it. But as with other cases in the philosophy of physics --- the enormous success of quantum theory in the face of the measurement problem, and of mainstream (non-algebraic) quantum field theory in the face of renormalisation --- it will not do simply to deny the coherence of a probability-based statistical mechanics given its empirical success.

But in any case, this mystery is largely dissolved by quantum mechanics. Though it is not often explicitly stated, the default assumption in foundations of statistical mechanics is that quantum and classical statistical mechanics are related as in Figure 1: mixed-state quantum mechanics is obtained from pure-state quantum mechanics via explicit addition of (mysterious) probabilities, just as Liouvillian classical mechanics is obtained from Hamiltonian classical mechanics, and the quantum/classical transition is from pure-state quantum mechanics to Hamiltonian classical mechanics.

\begin{figure}[H]\label{figure1}\caption{Quantum and classical statistical mechanics: standard view}
\begin{center}
\begin{tikzcd}[column sep = 8.0em, row sep = 5.0em]
CM \arrow{r}{+ \mbox{ probability}} & CSM \\
QM \arrow{u}[description]{\mbox{classical limit}}\arrow{r}{+ \mbox{ probability}}  & QSM \\
\end{tikzcd}
\end{center}
\end{figure}
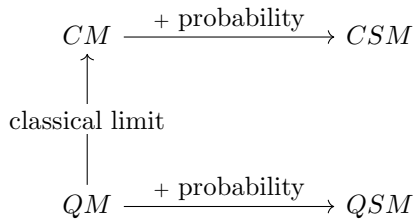
We have seen that this is not viable:
\begin{itemize}
\item The analogy between pure/mixed states and Hamiltonian/Liouvillian states is superficial: mixed states are already required as part of the formalism of quantum mechanics by subsystem considerations (section \ref{mixedstates}) and the addition of mysterious additional probabilities does not therefore generalise the dynamics in the quantum case as it does in the classical. Any equation of quantum mechanics in which probability distributions over (pure or mixed) states can always be reinterpreted as an equation about individual mixed states, so there is no inference from the empirical success of a given part of quantum statistical mechanics to the need for statistical-mechanical probabilities.
\item Study of the quantum-classical transition shows that quantum states (pure or mixed) correspond to phase-space distributions, not to phase-space points, and so argues for an understanding of that transition as a transition from quantum mechanics to Liouvillian, not Hamiltonian, classical mechanics.
\item Elementary considerations of wave-packet spreading, applied to dilute-gas systems with realistic parameters, reveals that Hamiltonian mechanics completely fails for such systems: inter-particle collisions are more like plane-wave interactions than classical hard-sphere scattering, and lead rapidly to a highly entangled state. So interpretation of the Boltzmann equation as a one-particle-marginal probability distribution (or relative-frequency description of) the trajectories of well-localised, deterministically evolving point particles is unjustified in our quantum universe. On the other hand, the derivation of that equation from Liouvillian dynamics via the BBGKY hierarchy remains valid even when the Liouvillian dynamics are understood as an approximate description of Schr\"{o}dinger dynamics, and indeed can be replicated exactly, \emph{mutatis mutandis}, in the Wigner-function formalism.
\item In a stochastic dynamics like the Brownian motion described by the Langevin equation, once we consider it quantum-mechanically we find that the stochasticity is itself quantum-mechanical, and persists even if the system and its environment begins in a pure state.
\end{itemize}
There is, in short, abundant evidence that `statistical-mechanical' probabilities are just quantum-mechanical probabilities. Considerations of the structure of quantum theory, of phase-space descriptions of quantum mechanics, and of specific models, all tell us that Liouvillian states need to be understood as quantum states in the regime where the Liouville equation approximates the Schr\"{o}dinger equation, and not as application of a new notion of probability to Hamiltonian dynamics. Hamiltonian dynamics is an approximation to Liouvillian dynamics in certain regimes, not its underpinning.\footnote{Of course, this is entirely compatible with the use of Hamiltonian dynamics as a formal method to analyse Liouvillian dynamics.} The correct relation between classical and quantum statistical mechanics is given by Figure 2.

\begin{figure}[H]\label{figure2}\caption{Quantum and classical statistical mechanics: improved view}
\begin{center}
\begin{tikzcd}[column sep = 4.0em,row sep=5.0em]
CM \arrow[leftarrow]{r}{\mbox{large-mass}}[swap]{\mbox{non-chaotic regime}} & CSM \\
  \mbox{pure-state } QM \arrow[phantom]{r}{\subseteq} & \mbox{mixed-state } QM \arrow{u}[description]{\mbox{decoherent regime}}\\
\end{tikzcd}
\end{center}
\end{figure}
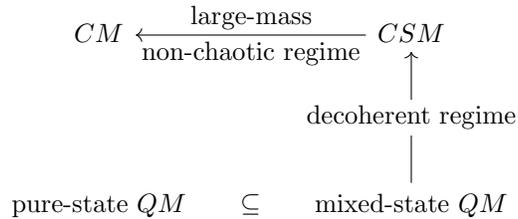

So in quantum statistical mechanics, one of the two puzzles of non-equilibrium statistical mechanics --- probability --- is resolved, or at any rate reduced to the problem of understanding quantum probabilities. The other puzzle --- the status of the assumptions underpinning irreversibility --- translates over to the quantum context \emph{mutatis mutandis}; a fuller understanding of them must await another day.

\section*{Acknowledgements}

Thanks to David Albert, Katherine Brading, Shelly Goldstein, Owen Maroney, Tim Maudlin, Wayne Myrvold, Simon Saunders, Chris Timpson, and especially Harvey Brown, for useful discussions.

\section*{Appendix}

Here I prove that if $\ket{\Omega}$ is an arbitrary state in NRQM with $M$ degrees of freedom, and that 
\be \ket{\vctr{q},\vctr{p}}\edf \op{T}_\vctr{p}\op{T}_\vctr{q}\ket{\Omega}\ee
is the state obtained by translating $\ket{\Omega}$ first by $\vctr{q}$ in position and then by $\vctr{p}$ in momentum, then 
\be 
\op{\Pi}\edf\frac{1}{(2\pi)^M}\int \dr{\vctr{q}}\dr{\vctr{p}}\proj{\vctr{q},\vctr{p}}{\vctr{q},\vctr{p}}=\id.
\ee
From this it follows that the continous family of operators $(2 \pi)^{-M}\proj{\vctr{q},\vctr{p}}{\vctr{q},\vctr{p}}$ is a POVM, or more rigorously that if phase space is decomposed into countably many disjoint Liouville-measureable sets $C_i$, then the family of operators
\be 
\op{\Pi}_i=\int_{C_i}\proj{\vctr{q},\vctr{p}}{\vctr{q},\vctr{p}}
\ee
is a POVM. In this section, $\hbar=1$.

To prove this, let $\ket{\psi}$ and $\ket{\varphi}$ be arbitrary Hilbert-space vectors.
 Inserting two complete sets of position eigenstates, we obtain
\be \label{app1}
\matel{\psi}{\Pi}{\varphi}=(2\pi)^{-M}\!\!
\int \dr{\vctr{q}}\dr{\vctr{p}}\dr{\vctr{x}}\dr{\vctr{y}}
\bk{\psi}{\vctr{x}}\bk{\vctr{x}}{\vctr{q},\vctr{p}}\bk{\vctr{q},\vctr{p}}{\vctr{y}}
\bk{\vctr{y}}{\varphi}.
\ee
Now, 
\be 
\bk{\vctr{x}}{\vctr{q},\vctr{p}}=
\bra{\vctr{x}}\!\op{T}_\vctr{p}\op{T}_\vctr{q}\!\ket{\Omega}=\e{i\vctr{x}\cdot \vctr{p}}\bra{\vctr{x}}\!\op{T}_\vctr{q}\!\ket{\Omega}
=\e{i\vctr{x}\cdot \vctr{p}}\bk{\vctr{x}\!+\!\vctr{q}}{\Omega}.
\ee
So we can rewrite (\ref{app1}) as
\be 
\matel{\psi}{\Pi}{\varphi}=(2\pi)^{-M}\!\!
\int \dr{\vctr{q}}\dr{\vctr{p}}\dr{\vctr{x}}\dr{\vctr{y}}
\bk{\psi}{\vctr{x}}\bk{\vctr{x}\!+\!\vctr{q}}{\Omega}\bk{\Omega}{\vctr{y}\!+\!\vctr{q}}
\bk{\vctr{y}}{\varphi}
\e{i(\vctr{x}-\vctr{y})\cdot \vctr{p}}.
\ee
The integral over $\vctr{p}$ can now be evaluated, and gives a delta function factor,
\be 
\int \dr{\vctr{p}}\e{i(\vctr{x}-\vctr{y})\cdot \vctr{p}}=(2\pi)^M \delta(\vctr{x}-\vctr{y}),
\ee
which we can in turn use to perform the integration over \vctr{y}, obtaining
\be 
\matel{\psi}{\Pi}{\varphi}=
\int \dr{\vctr{q}}\dr{\vctr{x}}
\bk{\psi}{\vctr{x}}\bk{\vctr{x}\!+\!\vctr{q}}{\Omega}\bk{\Omega}{\vctr{x}\!+\!\vctr{q}}
\bk{\vctr{x}}{\varphi}
\ee
After a change of variables $\vctr{z}=\vctr{x}+\vctr{q}$, this is just
\be 
\matel{\psi}{\Pi}{\varphi}=
\left(\int \dr{\vctr{x}}\bk{\psi}{\vctr{x}}\bk{\vctr{x}}{\varphi}\right)
\left(\int \dr{\vctr{z}}\bk{\vctr{z}}{\Omega}\bk{\Omega}{\vctr{z}}\right)
=\bk{\psi}{\varphi}\bk{\Omega}{\Omega}=\bk{\psi}{\varphi}.
\ee


\end{document}